\documentstyle[epsfig,12pt]{ioplppt}          
\begin{document}
\begin{flushright}
USC-99/006 , CITUSC/99-006\\
\hfill  hep-th/9911240
\end{flushright}

\title{Renormalization Group Flows from Five-Dimensional
Supergravity\ftnote{1}{ 
Talk presented at Strings '99 in Potsdam, Germany (July 19 -- 24, 1999).}}

\author{Nicholas P. Warner}

\address{Department of Physics and Astronomy,  
University of Southern California \\
Los Angeles, CA 90089-0484, USA}
\address{{\it and}}
\address{CIT-USC Center for Theoretical Physics \\
University of Southern California \\
Los Angeles, CA 90089-0484, USA}

%
%
%

\def\cN{{\cal N}}
\def\cR{{\cal R}}
\def\coeff#1#2{\relax{\textstyle {#1 \over #2}}\displaystyle}
\def\neql#1{{\cal N} \! =\! #1}  
\def\half{{1 \over 2}}
\def\IR{\relax{\rm I\kern-.18em R}}
\font\cmss=cmss10 \font\cmsss=cmss10 at 7pt
\def\ZZ{\relax\ifmmode\mathchoice
{\hbox{\cmss Z\kern-.4em Z}}{\hbox{\cmss Z\kern-.4em Z}}
{\lower.9pt\hbox{\cmsss Z\kern-.4em Z}}
{\lower1.2pt\hbox{\cmsss Z\kern-.4em Z}}\else{\cmss Z\kern-.4em
Z}\fi}
\font\ninerm=cmr9
\font\ninebf=cmbx9
\font\ninesl=cmsl9 
\def\nihil#1{{\it #1}}
\def\eprt#1{{\tt #1}}
 

\section{Introduction}

Perhaps the most extensively tested and studied version of the 
AdS/CFT correspondence  is that of $\neql{4}$ supersymmetric
Yang-Mills theory in $(3+1)$  dimensions and the compactification
of $IIB$ supergravity on $AdS^5 \times S^5$ \cite{JMalda}.  The relationship
between supergravity fields and Yang-Mills operators can most
easily be mapped out from the representation structure of the extended
superconformal algebra.  Taken at face
value one might be tempted to conclude that the 
AdS/CFT correspondence itself may be little more than 
a convenient way to encode this representation theory, but   
this view is too facile, and perhaps one  of the remarkable lessons
from the last year has been just how deeply this correspondence 
works, even far from the conformal regime.

One of the purposes of this talk is to describe one 
of the more non-trivial tests of the correspondence \cite{KPW,FGPWa}, 
a test in which one moves far from the  $\neql{4}$ superconformal regime of the 
Yang-Mills theory:  I will discuss how supergravity correctly
describes an $\neql{1}$ supersymmetric renormalization group flow,
and how this flow can be checked rather well against known
field theory results.  A by-product of this 
study of renormalization group flows is the discovery
of a general $c$-theorem for any such flow described via supergravity.
The $c$-function does everything that one wants of such a
function: it is monotonic decreasing on all flows to the
infra-red, and is only stationary at conformal fixed points, and
at such fixed points it is equal to the central charge of the
fixed-point theory.    On a more philosophical level,   given that 
the renormalization group flow is being modelled by a five-dimensional
cosmological solution, it is also satisfying that the central charge 
is related to the inverse of the cosmological entropy density. 

Using supergravity to describe renormalization group flows was 
first discussed in \cite{GPPZa,JDFZ}, but these flows were non-supersymmetric
and they went to unstable supergravity vacua.  The lack of supersymmetry
makes it hard to independently verify the results via field theory,
and the lack of stability suggests that the IR field theory limit may in fact
be pathological (non-unitary).  The solution and flow discussed here has neither of
these problems, and indeed the supergravity computations match perfectly with
independent field theory results.

A secondary purpose of this talk will
be to revisit some ancient issues in supergravity, most
particularly that of consistent truncation, and discuss
the relevance of these issues to the phase structure
and flows in Yang-Mills theory.   The final part of my talk will
focus on how  supergravity distinguishes between perturbations
of the Yang-Mills action and states in the Hilbert space.  As an 
illustration of this I will discuss the supergravity description of
part of the Coulomb branch of the  $\neql{4}$ supersymmetric
Yang-Mills theory \cite{FGPWb}.

\section{Gauged $\neql{8}$ Supergravity in Five Dimensions}

\subsection{Ancient Issues: Embedding and Consistent Truncation}

Maximal supergravities in any dimension can usually be obtained
by trivial dimensional reduction of the eleven dimensional supergravity 
on a torus.  Such supergravities are ``ungauged'' in that all the vector fields
have abelian gauge symmetries, and these gauge fields are not minimally
coupled to the fermions.  Gauging refers to converting   part of  the 
global symmetry into a local one by introducing the appropriate couplings
by hand, with a coupling constant, $g$, and then constructing the
supersymmetric model by the ``Noether procedure.''  If the gauge
group is non-abelian then one can show very generally that any ground state
that leaves the symmetries unbroken
is necessarily anti-de Sitter space, with a cosmological constant 
proportional to $g^2$.   There are sometimes several options  as
to the choice of gauge group, but here I will only consider the
gauged theories with maximal supersymmetry and the maximal $SO(M)$ 
gauge symmetry.   Such theories  are related to sphere compactification
of IIB and eleven dimensional supergravity.

It is important to stress that gauged supergravity theories are consistent, 
classical field theories in their own right.  Their structure is generically very
non-linear, particularly in the scalar sector.  However, it is also
important to remember that the complete structure of these theories
is completely determined by: (i) Supersymmetry, gauge and global symmetries, and
(ii) The perturbative field content.  Given this, and the assumption of
only second-derivative actions, the Noether procedure is essentially 
unambiguous.

It is a separate issue as to how gauged supergravities may be related to
compactifications of higher dimensional supergravities, and indeed,
one of the key issues is  whether they are actually {\it embedded} in the
higher dimensional theory.  Since the lower dimensional theory is a 
truncation of the higher dimensional theory down to some subset of
``lowest modes'' there is the obvious question as to whether the truncation
is consistent with the equations of motion of the higher dimensional field
theory, or whether the lower dimensional field theory is merely some
effective low energy truncation.   In this context, ``embedding'' 
 is the strongest option: it means that the equations
of motion of the higher dimensional field theory are exactly satisfied
if one satisfies the equations of motion of the lower dimensional theory.
In particular, a solution to the lower dimensional theory automatically
yields a solution to the higher dimensional theory.

Embedding, or consistent truncation is trivial for toroidal compactifications
where one takes all fields to be independent of the toroidal coordinates, but it
is highly non-trivial for the sphere compactifications that lead to maximal
gauged supergravities.  Embedding of gauged supergravity is almost completely proven 
for the  $S^7$ compactification  and $S^4$ compactification of eleven dimensional 
supergravity \cite{ConTrunc}.  It has not been proven for gauged $\cN = 8$ supergravity 
in five  dimensions and the $S^5$ compactification of the IIB theory.  
However there is a significant body of evidence to support this
correspondence, and the similarities with the other sphere compactifications 
make the case very convincing\footnote{Consistent truncation has
been recently proven for a subsector of the theory 
\cite{HPT}.}.  I will therefore assume that the five-dimensional
gauged supergravity is indeed embedded in the IIB theory, and proceed.

\subsection{The Field/Operator Correspondence}

The field content of gauged $\neql{8}$ supergravity is a graviton, eight
spin-${3 \over 2}$ fields, twelve tensor gauge fields,   fifteen vector 
gauge fields, forty-eight spin-$\half$ fields, and forty two scalars.
The ungauged theory has a global $E_{6(6)}$ symmetry and a composite local $USp(8)$
symmetry, with the scalars parametrized in terms of the coset $E_{6(6)} \over
USp(8)$.  In the  gauged supergravity theory \cite{GaugSG} the global $E_{6(6)}$ 
is replaced by a local $SO(6)$ symmetry and a global $SL(2,\IR)$.
The latter can be thought of as the $SL(2,\IR)$ symmetry inherited from the
IIB theory, with the dilaton and axion corresponding to the coset  
$SL(2,\IR) \over U(1)$.  At the quantum level this $SL(2,\IR)$ must therefore be
broken to $SL(2,\ZZ)$.  

Even though the $E_{6(6)}$ symmetry is broken in the gauged theory
it is still convenient to think of the scalar fields as living on  
$E_{6(6)} \over USp(8)$, but now the scalars
also have a potential $\cal V(\phi)$ proportional to $g^2$.  This potential
is invariant under $SO(6) \times SL(2,\IR)$.
 
The AdS/CFT correspondence implies \cite{JMalda,GKP,EWhol} that this $\neql{8}$ 
gauged  super\-gravity multiplet must correspond to the $\neql{4}$ Yang-Mills
energy-momentum tensor supermultiplet, with the local $SO(6)$ 
symmetry dual to the $\cal R$-symmetry.   The scalar fields of the 
supergravity lie in the following $SO(6)$ representations:
${\bf  1 \oplus 1  \oplus 10 \oplus \overline{10}  \oplus 20'}$, with the 
two singlets being the ten-dimensional dilaton and axion.    From this
one can deduce that these supergravity scalars correspond, 
respectively, to the gauge coupling, the $\theta$-angle,
the fermion bilinear operators and the scalar
bilinear operator  of the Yang-Mills
theory.  The latter have the form:
\begin{equation}
{\rm Tr}~(\lambda^i ~\lambda^j)\ ,  \quad
{\rm Tr}~(\bar \lambda^{\bar i} ~\bar \lambda^{\bar j}) \ , \quad
{\rm Tr}~(X^A ~X^B)  ~-~ {1 \over 6} ~\delta^{AB}~{\rm Tr}~(X^C ~X^C) \ . 
\label{massterms}
\end{equation}

\subsection{Phases and operator product algebras}

When one combines the embedding of gauged supergravity
with the AdS/CFT correspondence one is led to some interesting
statements about the Yang-Mills theory.  Consistent truncation implies
that  the operator  product of fields in the energy-momentum supermultiplet
must be closed at large $N$, a fact that has been demonstrated in 
\cite{OPEfourD}.   Conversely, the fact that
this operator product algebra does not close at finite $N$ implies that
gauged $\neql{8}$ supergravity is an incomplete quantum theory. 
The presence and exact knowledge of the supergravity potential  
(and other highly non-linear structure in supergravity) implies a
completely determined, highly non-trivial operator algebra.
Furthermore, critical points of the potential give rise to 
anti-de Sitter vacua in supergravity, which will, under appropriate
conditions, imply non-trivial conformal fixed points for
the Yang-Mills theory.  The supergravity scalars whose VEV's lead 
to the new critical point tell one exactly what relevant operators in the
Yang-Mills theory would drive a flow to this fixed point in the infra-red.
Moreover, constructing the 
supergravity kink solutions that interpolate between these AdS 
ground states should be equivalent to constructing explicitly the 
renormalization group flow \cite{GPPZa,JDFZ,FGPWa}.   
Thus, in a very real sense, the 
supergravity potential should map out the phase diagram of the 
$\neql{4}$ Yang-Mills theory under mass perturbations.

In \cite{KPW} the  critical points of supergravity potential with at least
$SU(2)$ invariance were classified. The results are summarized in  Table 1.    
{}From the work of \cite{MHKS} one can deduce that the central charge at the new 
infra-red fixed  point is related  to the central charge at the UV fixed 
point according to:
\begin{equation}
 c_{IR}~=~ \bigg( {\Lambda_{IR} \over \Lambda_{UV}} \bigg)^{-3/2}~ c_{UV} \ , 
\label{centch}
\end{equation}
where $\Lambda$ denotes the cosmological constant of the corresponding 
five-dimensional supergravity solution.

To be a stable solution of the supergravity theory the solution must 
either be supersymmetric \cite{GHW} or the small oscillations must satisfy 
the Breitenlohner-Freedman (BF) condition \cite{PBDF}.  If there are 
supergravity modes that fail the BF-condition then the corresponding
Yang-Mills operators seem to have complex conformal dimension, violating
unitarity, and so corresponding phases appear to be pathological.   
The stability  analysis for the $SO(5)$ point was done in \cite{JDFZ}, 
and the instability of the other non-supersymmetric points was established 
by Pilch \cite{KPpriv}.

It thus seems that only the last critical point represents a viable,
non-trivial phase of large $N$ Yang-Mills theory.  It is, of course 
natural to ask whether this should also represent a phase at finite $N$, and
the AdS/CFT correspondence implies that this question is equivalent to asking
whether the supergravity solution is a good vacuum for the IIB string.
The general dogma in this area states that this is true for $\neql{4}$ 
supersymmetric supergravity solutions, and suggests that there should be
a string vacuum ``nearby'' for $\neql{2}$ supersymmetric supergravity solutions.
Thus finding a corresponding Yang-Mills phase represents a very non-trivial, 
non-linear test of the AdS/CFT correspondence (and of the general
dogma about supersymmetric solutions).

\goodbreak

{\vbox{ {
$$
\vbox{\offinterlineskip\tabskip=0pt
\halign{\strut\vrule#
&~$#$~\hfil\vrule
&~$#$~\hfil\vrule
&~$#$~\hfil\vrule
&~$#$~\hfil\vrule
&~$#$\hfil
&\vrule#
\cr
\noalign{\hrule}
&
{\rm Unbroken\ Gauge}
&
{\rm Cosmological}
&
{\rm Unbroken}
&
{\rm Central Charge}
&
{\rm Stable?}
&\cr
&
{\rm Symmetry}
&
{\rm Constant}
&
{\rm Supersymmetry}
&
{\rm c}_{IR}/{\rm c}_{UV}
&
&\cr
\noalign{\hrule}
&
SO(6)
&
-{3 \over 4} g^2
&
\cN = 8
&
1
&
{\bf Yes}
&\cr
&
SO(5)
&
- {3^{5/3} \over 8} g^2
&
\cN=0
&
{2 \sqrt{2} \over 3}  
&
{\bf No}
&\cr
&
SU(3)
&
-{27 \over 32} g^2
&
\cN =0
&
{16 \sqrt{2} \over 27}  
&
{\bf No}
&\cr
&
SU(2) \times U(1) \times U(1)
&
- {3 \over 8}({25 \over 2})^{1/3} g^2
&
\cN=0
&
{4 \over 5}   
&
{\bf No}
&\cr
&
SU(2) \times U(1)
&
-{2^{4/3} \over 3}g^2
&
\cN =2
&
{27 \over 32}  
&
{\bf Yes}
&\cr
\noalign{\hrule}}
\hrule}$$
\vskip-10pt
\noindent{\bf Table 1:}
{\sl
The $SU(2)$ invariant critical points of the $\neql{8}$ supergravity 
potential.   } \vskip10pt}}}

{}From the supergravity scalar vevs at this non-trivial, supersymmetric critical 
point one finds that in the $\neql{4}$ Yang-Mills theory one is turning on a 
mass for one of the four fermions, and two of the six scalars.  Indeed, writing
everything in terms of $\neql{1}$ superfields, one sees that this new phase 
arises from giving a mass to exactly one of the three
adjoint hypermultiplets.  Based upon this, several authors \cite{LSflow,FGPWa} 
have identified this infra-red phase as precisely one of those discovered
by Leigh and Strassler in 1995 \cite{RLMS}.  The global symmetries, supersymmetry,
the UV end of the flow, the central charge and the IR anomalous dimensions 
have all been checked, and the match is perfect.  Of course, once one has
established the existence of such a phase, a number of the properties, like
anomalous dimensions, follow as a consequence of the  $\neql{1}$  
superconformal symmetry.  However,  the remarkable thing is 
that supergravity knows about the existence of the fixed point, and
that it requires the extremely non-linear structure of the potential
to see it.  Conversely, had supergravity not found this fixed point,
it would have cast significant doubt on the ability of AdS/CFT to go
beyond the most basic, second order perturbative details.

\section{Renormalization Group Flows}

\subsection{The $\neql{1}$ flow}

Consider now the supergravity scalars $\varphi_1, \varphi_2$, that
are dual to the mass parameters of a fermion and the two scalar fields 
that constitute an $\neql{1}$ Yang-Mills hypermultiplet.  On this subsector 
one can write the supergravity potential in the canonical form:
\begin{equation}
V ~=~ {g^2 \over 8}~\sum_{j = 1}^3 ~\bigg| {\partial W
\over \partial \varphi_j} \bigg|^2 ~-~ {g^2 \over 3}~\big|W \big|^2 \ ,
\label{VfromW}
\end{equation}
where
\begin{equation}
W~=~ {1 \over 4 \rho^2}~ \Big[\cosh(2 \varphi_1)~
( \rho^{6}~-~ 2)~ - ( 3\rho^{6} ~+~ 2 ) \Big] \ , 
\label{Wreduced}
\end{equation}
and $\rho \equiv e^{{1 \over \sqrt{6}} \varphi_2}$.  The scalar 
kinetic term has the usual form: ${1 \over 2}  \sum_j  g^{\mu \nu}  
(\partial_\mu \varphi_j) (\partial_\nu \varphi_j)$.

The $\neql{8}$ supersymmetric critical point is at
$\varphi_j \equiv 0$, and there is a $\ZZ_2$
equivalent pair of  $\neql{2}$ supersymmetric critical points at:
\begin{equation}
\varphi_1 ~=~ \pm \coeff{1}{2}~\log(3) \ ,   
\quad \varphi_2   ~=~ \coeff{1}{\sqrt{6}} \log(2)   \ .
\label{susyGS}
\end{equation}

To construct the kink corresponding to the renormalization group 
flow one requires the  five-metric to take the form:
\begin{equation}
ds^2 = e^{2 A(r)} \eta_{\mu\nu} dx^\mu dx^\nu - dr^2 \ .
\label{genmetric}
\end{equation}
(I will adopt the conventions of \cite{FGPWa} throughout.)   
As $r \to \infty$ the scalars, $\varphi_j$, must vanish
and one must have $A(r) \sim const. ~r$ so that the background is 
asymptotic to the $\neql{8}$ supersymmetric AdS background 
at infinity.  At the other end of the flow ($r \to - \infty$),
the scalars must approach either of the critical values (\ref{susyGS}),
and once again $A(r) \sim const. ~|r|$ as one approaches
the new conformal fixed point.  

Since the flow is $\neql{1}$ supersymmetric one should be able to
find a five-dimensional kink that preserves this amount of supersymmetry.
The equations of motion for such a flow can be obtained by finding
solutions to the vanishing of the supersymmetry variations of the
supergravity fermions: $\delta \psi_\mu^a = 0$, $\delta \chi_{abc} = 0$.
{}From this one can show that there is a supersymmetric flow if and only if:
\begin{equation}
{d \varphi_j \over d r} ~=~ {g \over 2}~{\partial W \over \partial 
\varphi_j} \ , \qquad {\rm and} \qquad A'(r) ~=~ - {g \over 3}~W \ .
\label{susyflows}
\end{equation}
These equations mean  that the flow is determined by the steepest 
descent of the superpotential, and that the cosmology ($A(r)$)
is determined directly from this steepest descent. 

Figure 1 shows
the contour maps of the potential $V$, and the superpotential, $W$.
These functions have a $\ZZ_2$ symmetry under $\varphi_1 \to -\varphi_1$,
and thus there are two equivalent copies of each non-trivial critical point.
The contour map of $V$ shows five critical points corresponding to
the first, third and fifth entries in Table 1.  Critical points
of $W$ yield supersymmetric vacua, and Figure 1 also shows these
as well as a steepest descent path.  Note that the steepest descent path
has a parabolic form near the central critical point:
\begin{equation}
\varphi_1 \sim a_0~e^{- r} \ ,\qquad \varphi_2 \sim \sqrt{8 \over 3}~a_0^2 ~
r e^{- 2r} + a_1 ~e^{- 2r} \ , \quad {\rm as} \ \ r \to \infty\ .
\label{massasymp}
\end{equation}
This is, of course, precisely what one needs for the Yang-Mills scalar masses
to be the square of the Yang-Mills fermion mass along the supersymmetric flow.
%

\begin{figure}
\centerline{\epsfig{figure=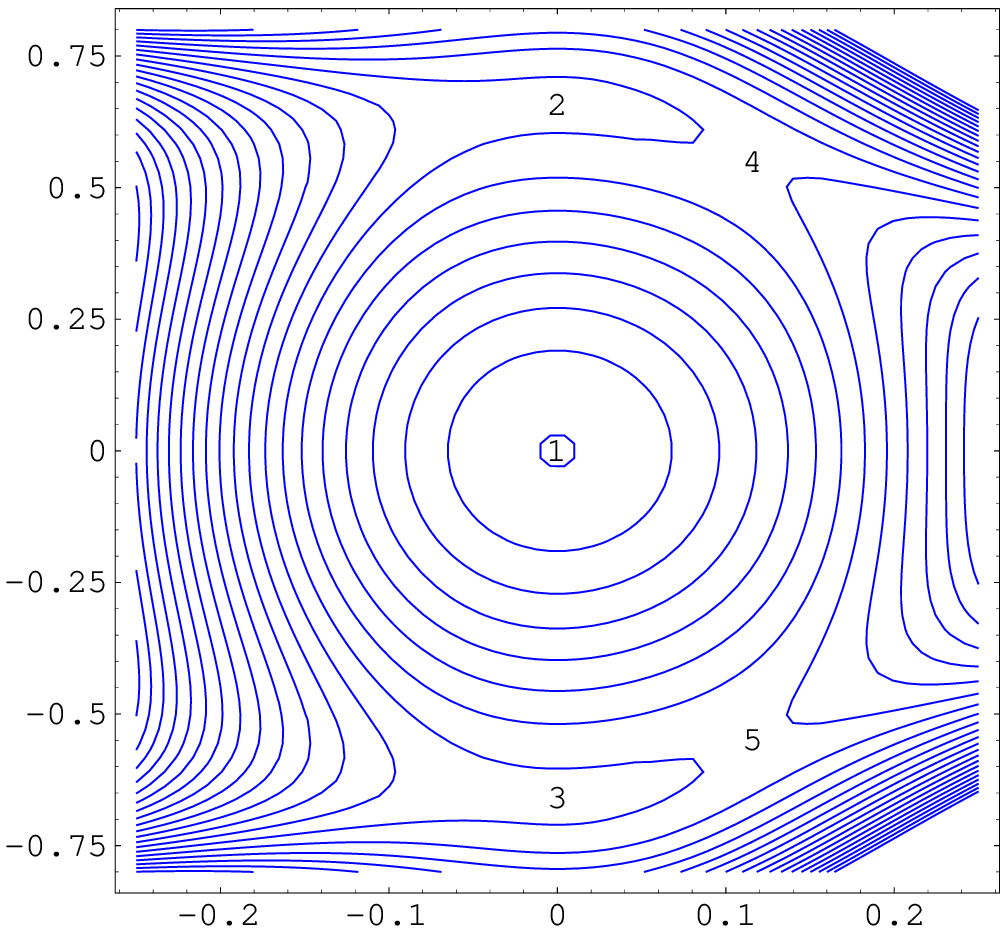,width=2.5in}
\hskip 0.25in \epsfig{figure=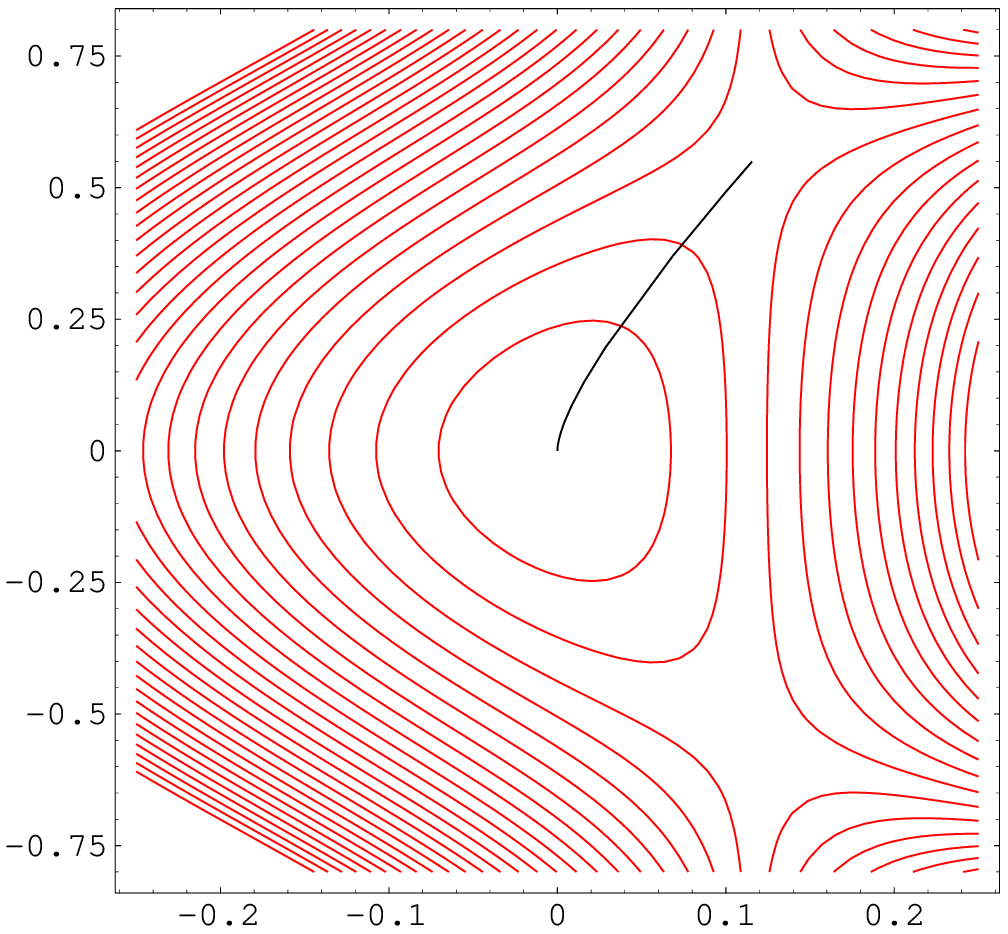,width=2.5in}  }
\leftskip 2pc
\rightskip 2pc\noindent{\ninesl  \baselineskip=8pt  
{\ninebf Fig.~1}: The contour map of $V$ (on the left) and $W$ (on the right), 
with $\varphi_1$ on the vertical axis and $ {1 \over \sqrt{6}}\varphi_2  $
on the horizontal axis.   The five labelled points are the only extrema of
$V$ in this plane.  The points 1, 4 and 5 are supersymmetric and so show
up as critical points of $W$.  The points 2, 3 are $\ZZ_2$ equivalent
copies of the  SU(3)-invariant critical point.  A numerical solution of the 
steepest descent equations is shown superimposed on the contour plot of $W$.} 
\end{figure}

%
\subsection{The $c$-Function}

If one wishes to describe a general renormalization
group flow in terms of holography then one considers 
a five-metric of the form (\ref{genmetric}) with some five-dimensional
matter on the brane.  It is relatively elementary to show \cite{FGPWa} 
that if the energy-momentum tensor of this matter satisfies the null
energy condition then that the following function is a $c$-function 
for the flow:
\begin{equation}
{\cal C}(r) = {{\cal C}_0 \over (A'(r))^{3}} \ , 
\label{CFunctionGen}
\end{equation}
where ${\cal C}_0$ is a suitably chosen constant. 

Specifically, ${\cal C}(r)$ is monotonic decreasing along renormalization
group flows to the infra-red.  It is only stationary at conformal fixed points
of the flow, and by proper choice of ${\cal C}_0$ the function ${\cal C}(r)$
is equal to the central charge of the conformal theory at any fixed point.
For the  flows of $\neql{4}$ Yang-Mills considered earlier, one sets 
 ${\cal C}_0 =g^3 N^2 / 32 $, which gives ${\cal C}=a=c={N^2 \over 4}$ at
the $\neql{4}$ supersymmetric point.   The same $c$-function and a 
restricted version of the $c$-theorem was also  obtained in  \cite{GPPZa}. 

Interestingly enough, for the supersymmetric flows one can use 
(\ref{susyflows}) to obtain:
\begin{equation}
{\cal C}(r) = -  {27 \over g^3}~ {{\cal C}_0 \over W^{3}} \ , 
\label{explicitCFunctionGen}
\end{equation}
where $W$ is the superpotential.  This form of the $c$-function is
much more in the spirit of Zamolodchikov's  $(1+1)$-dimensional 
$c$-function in that it is explicitly written as a function of the 
couplings, whereas (\ref{CFunctionGen}) only has an implicit 
dependence on the couplings.  It is also interesting to note that
the renormalization group flows (\ref{susyflows}) also represent 
steepest descents of the $c$-function (\ref{explicitCFunctionGen}).

\section{Flows to {\it Hades} and the Coulomb Branch}

In the previous section I considered a very specific flow
that follows a ridge line of the superpotential (\ref{Wreduced}) to the 
non-trivial fixed point.  However, {\it any} solution to (\ref{susyflows})
represents a supersymmetric solution to the five-dimensional supergravity
theory: indeed  Figure 2 shows a family of supersymmetric flows 
superimposed on the contours of $W$.  It is evident that all the  
flows (other than the ridge-line flow) go to infinite values
of the scalar fields, and thus to infinite values of $W$ (called {\it Hades}
by one of my illustrious collaborators).  Since these flows are part of
the supergravity theory, and  have some interpretation in terms
of perturbations of the gauge theory at its UV fixed point, it would seem
that all these flows should also have some gauge theoretic interpretation.
However instead of tackling this problem directly, I will consider
a larger, but simpler family of flows to Hades; a family that includes
the flows along the horizontal axis in Figure 2.

\begin{figure}
\centerline{\epsfig{figure=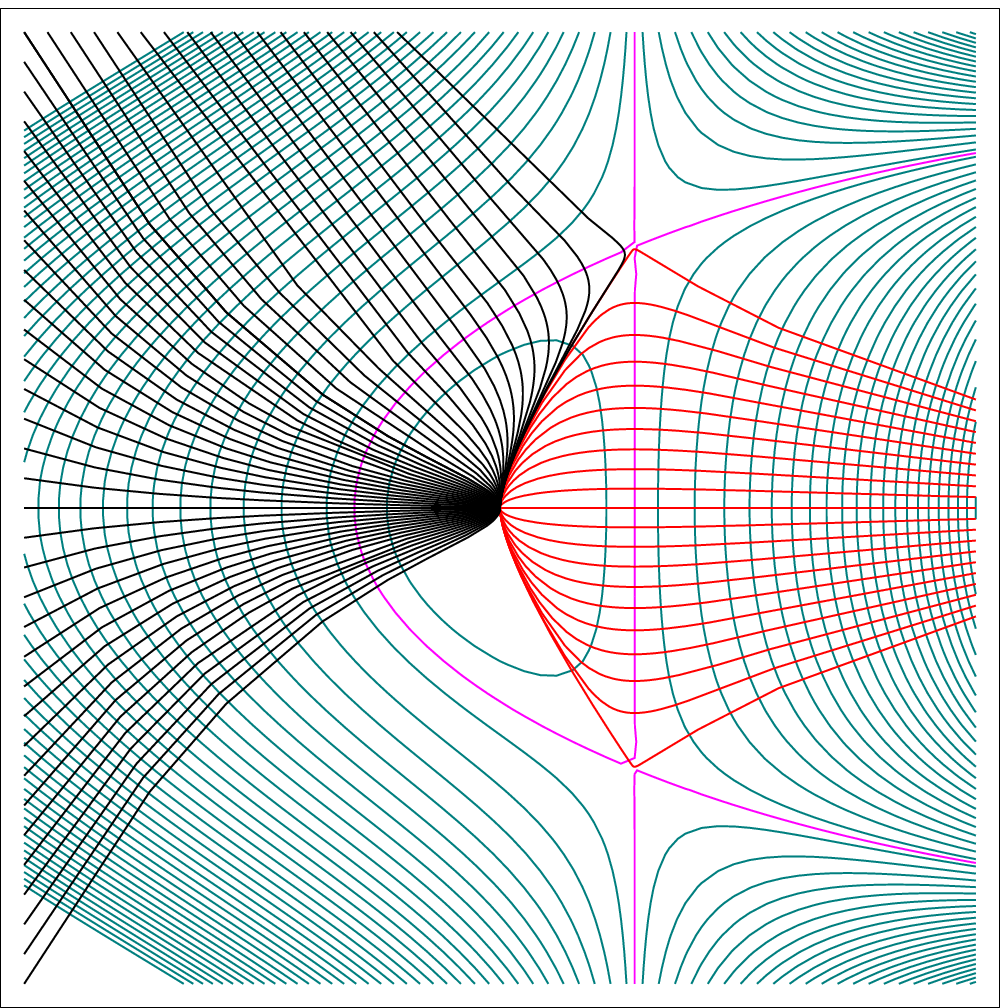,width=2.8in}  }
\leftskip 2pc
\rightskip 2pc\noindent{\ninesl \baselineskip=8pt  
{\ninebf Fig.~2}: The contour map of  the superpotential  (\ref{Wreduced})
showing families of steepest descent paths from the central critical point. 
Each such path represents an ${\cal N}\! =\! 1 $ supersymmetric solution.} 
\end{figure}

Consider all the supergravity scalars in the ${\bf 20'}$ of $SO(6)$. 
These are naturally parametrized by a matrix $S \in SL(6,\IR)/SO(6)$,
and on this subsector the potential and superpotential have the form:
\begin{equation}
 V~=~ -{g^2 \over 32}~\Big[~ ({\rm Tr}(M))^2 - 2 {\rm Tr} (M^2)~ \Big] \ , \qquad
 W ~=~ - {1\over 4}~{\rm Tr} (M) \ ,
\end{equation}
where $M\equiv S S^T $.  Using the $SO(6)$ symmetry one can take
$M$ to have the form $M = {\rm diag}(e^{2 \beta_1},e^{2 \beta_2},e^{2 \beta_3},
e^{2 \beta_4},e^{2 \beta_5},e^{2 \beta_6})$, with $\sum_a \beta_a =0$.  
If $\varphi_j$ are  orthonormal linear combinations of the $\beta_a$, then
once again one obtains supersymmetric solutions to the supergravity
theory if one satisfies (\ref{susyflows}).  However, these solutions
preserve $\neql{4}$ supersymmetry on the brane.
One can also show that while the flows near the UV fixed point are
generically complicated, far from the UV fixed point the flows 
settle down to one of five classes:
\begin{itemize}
\item {Two $SO(5)$ invariant flows defined by  $\vec \beta ~ =~  
\pm (1,1,1,1,1,-5) \mu$ }
\item {Two $SO(4)\times SO(2)$ invariant flows defined by  $\vec \beta ~ =~  
\pm (1,1,1,1,-2,-2) \mu$ }
\item {An $SO(3)\times SO(3)$ invariant flow  defined by  $\vec \beta ~ =~  
 (1,1,1,-1,-1,-1) \mu$ }
\end{itemize}
where $\mu$ is a parameter.  
The corresponding five-metrics are regular for the $SO(5)$ and 
$SO(4)\times SO(2)$ invariant flows with $\mu>0$ and
the ``$+$'' choice.  The regularity, and physical interpretation
 of the corresponding ten-dimensional metrics will be discussed
in the talk by Steve Gubser.

{}From the Yang-Mills perspective, the scalars in the ${\bf 20'}$ of $SO(6)$
are dual to the last set of operators in (\ref{massterms}).  However these 
supergravity scalars cannot correspond to turning on masses since
the flows described above all preserve the full $\neql{4}$ supersymmetry.
These supergravity scalars must therefore represent moduli of
the $\neql{4}$ theory, and indeed these flows must be 
along the Coulomb branch.   One can see this fact 
directly by constructing the corresponding ten-dimensional metrics, and looking
at the distribution of branes that make up the sources  (see \cite{FGPWb}, and 
the talk by  S.~Gubser).   
More generally, this situation raises the issue of  when a supergravity scalar 
represents a mass term or a field theory modulus.  The resolution of this was 
given in \cite{EWhol,BKL}, and there is a direct characterization that can be 
expressed entirely from the five dimensional perspective.

Recall that a general (non-supersymmetric) flow in the supergravity
scalar sector is determined by a second order differential equation:
\begin{equation}
{d^2 \varphi_j \over d r^2} ~+~ 4 A'(r)~{d \varphi_j \over d r } ~=~
{\partial V \over \partial \varphi_j}\ . 
\end{equation}
Near the UV critical point this equation has a general solution with asymptotic
behaviour:
\begin{equation}
 \varphi_j  ~\sim~  a_j~r e^{-2r} ~+~ b_j~e^{-2r}\ . 
\label{genasymp}
\end{equation}
The first of these modes is non-normalizable on the $AdS_5$, and so 
couples non-trivially to operators on the boundary, and thus represents
a ``mass insertion'' on the brane.  The second mode represents a 
normalizable wave-function on $AdS_5$, and thus must represent a state
in the Hilbert space, which here means a modulus of the Coulomb branch
of the $\neql{4}$ theory.  Thus a general flow can involve a mixture 
of mass terms and vevs.  The supersymmetric flows are determined
by first order equations (\ref{susyflows}), and they select a 
unique solution, and one can easily check that the flows considered
in this section involve {\it only} normalizable modes ($a_j =0$ in
(\ref{genasymp})), whereas the $\neql{1}$ flow discussed earlier has
asymptotics (\ref{massasymp}), which manifestly involves non-normalizable
modes.

Thus even the apparently pathological {\it flows to Hades} in
the five-dimensional supergravity have a sensible physical interpretation
on the brane.  Moreover the five-dimensional supergravity also
captures the subtleties of operators and states in the AdS/CFT 
correspondence.

\section{Final Comments}

As outlined in the introduction, the primary purpose here
was to describe some extremely non-trivial (and successful) tests of the 
AdS/CFT correspondence, and its generalization to renormalization
group flows.  I also wished to
highlight the utility of five-dimensional gauged supergravities
as a tool for examining the non-linear structure of  
field theories on the brane, and to show how some ancient issues
of gauged supergravity theories have new relevance to holographic 
field theories.  I would like to conclude in a similar spirit by raising
some related, but as yet unresolved issues.  

First, in the work that led up to \cite{FGPWa,FGPWb} we also found a broad 
family of  $\neql{1}$ and $\neql{2}$ flows to Hades.  We are still looking at
some of the details, but these flows certainly
represent combinations of massive and Coulomb branch flows.   

It is an empirical fact that  supersymmetric flows appear
to be governed by steepest descent of a superpotential, $W$, and  
hence of the $c$-function.  It is not clear how far this can be 
generalized.  Specifically, we have found a $c$-function for general 
flows, but it is not written directly as a function of the couplings, and it
is not known whether a general RG flow is necessarily a steepest 
descent of this $c$-function.  However, the recent results of \cite{DFGK} 
suggest that some non-supersymmetric generalizations are indeed possible.

Returning to ancient supergravity issues:  there are 
non-supersymmetric critical points, and all the known ones 
are unstable.  Is this true in general, {\it i.e.} does stability
also imply supersymmetry?  Additionally, what is the significance of 
the unstable critical points for the field theory on the
brane?  Are they merely some large $N$ pathology?  

It would obviously be interesting to determine  all critical points 
of the potential, or find some way of controlling the problem.
For black hole solutions based on Calabi-Yau compactifications one finds 
the ``flow'' to the horizon induces a potential on the Calabi-Yau moduli
space, and that that potential selects certain natural arithmetic points
as attractors for the flow \cite{Attractors}.   It would be interesting if there 
were some arithmetic principle in $E_{6(6)}$ for classifying the
critical points of the supergravity potential.

One last issue:  it would be very interesting to find supersymmetric flows
similar to those discussed here, but in which the dilaton runs.
Most particularly it would be nice to do this with the recently
discovered $\neql{1}$ flows of \cite{GPPZb}, and thereby obtain a flow
to pure $\neql{1}$ gauge theory  with a running coupling.
 
\ack

I would like to thank my collaborators,
Dan Freedman, Steve Gubser and Krzysztof Pilch. I
am also very grateful to the organizers of Strings `99 for
opportunity to present this material, and to the Theory Division
at CERN for its hospitality during the major part of this
research.  This work was supported in part  
by funds provided by the DOE under grant number DE-FG03-84ER-40168.

\end{document}